# Methodology to assess prosumer participation in European electricity markets

Rubén Rodríguez-Vilches[1]*, Francisco Martín-Martínez[1], Álvaro Sánchez-Miralles[1], Javier Rodrigo Gutiérrez de la Cámara[2], Sergio Muñoz Delgado[2]

[1] Universidad Pontificia Comillas. Escuela Técnica Superior de Ingeniería ICAI.
Instituto de Investigación Tecnológica.
Santa Cruz de Marcenado 26, 28015, Madrid, Spain

[2] OMI, Polo Español S.A. (OMIE)
C/ Alfonso XI, nº 6, 28014 Madrid, Spain

The emergence of distributed generation and the electrification of demand have opened the possibility for prosumers to participate in electricity markets, receiving economic benefits on their bills and contributing to the reduction of carbon emissions, aligning with United Nations Sustainable Development Goal 7. Consumers and prosumers can participate through implicit and explicit demand flexibility and (collective) self-consumption. This study analyses the potential markets in which prosumers can participate and indicates whether these are currently open. The markets studied include day-ahead, intraday, ancillary services, adequacy services, constraint management, and local flexibility markets. Additionally, collective self-consumption is analysed as a service through which prosumers can participate in the electricity market. Previous studies are usually focused on a single market or in a single country, making impossible a complete comparison. This analysis has been done in Spain, Italy, Croatia, and the United Kingdom as representative countries to obtain a methodology to assess countries' openness to prosumer participation in electricity markets, comparing regulatory frameworks and assigning scores based on their prosumer inclusion across various markets. This work updates current literature reviews with the changes and a new description of local market designs in Spain. This methodology can be used to compare other countries' grade of openness. The results of this study show that the analysed countries can be categorised into three groups: almost open, partially open, and closed markets. Analysing the differences, recommendations on the following steps to foster user participation are suggested for each group.

*Corresponding author

Email addresses: rrodriguezv@comillas.edu (R. Rodríguez-Vilches) fmartin@comillas.edu (F. Martín-Martínez), alvaro@comillas.edu (A. Sánchez-Miralles),  jrodrigo@omie.es ( J.R. Gutiérrez de la Cámara), smunoz@omie.es (S. Muñoz Delgado).

*Abstract* — The emergence of Distributed Generation and the electrification of demand have allowed prosumers to participate in electricity markets and receive economic benefits on their electricity bills. Prosumers can participate through implicit demand flexibility (through prices in electricity tariffs), explicit demand flexibility (through market signals to modify demand generating savings or incomes) and (collective) self-consumption. This study analyses the potential markets in which prosumers can participate and indicates whether these are currently open. The markets studied include day-ahead, intraday, ancillary services, adequacy services, constraint management, and local flexibility markets, with Spain, Italy, Croatia, and the United Kingdom as the focus countries. Additionally, collective self-consumption is analysed as a service through which prosumers can participate in the electricity market. Based on the commonalities found in the studied countries, a methodology is proposed to determine the openness of any other country. The results show that the analysed countries can be categorised into three groups: almost open, partially open, and closed markets.

*Highlights*

- *This paper proposes different indicators to compare the openness of prosumer participation in electricity markets in a straightforward manner.*

- *It is analysed the following markets: day-ahead, intraday, ancillary services, adequacy services, constraint management, and local flexibility markets.*

- *The proposed indicator has been used to assess the openness of electricity markets in Spain, Italy, Croatia and the United Kingdom.*

- *Results suggest that most countries still do not offer significant opportunities for prosumer participation in the countries analysed.*

*Word Count*: 9,915

*List of abbreviations*

| | |
|---|---|
| aFRR | Automatic Frequency Restoration Reserve |
| ARERA | Autorità di Regolazione per Energia Reti e Ambiente (National Regulation Agency in Italy) |
| BM | Balancing Mechanism |
| BRP | Balancing Responsible Party |
| BSP | Balancing Service Provider |
| CSC | Collective Self-Consumption |
| D | Delivery day |
| DA | Day-Ahead market |
| DC | Dynamic containment |

| | | |
|---|---|---|
| DER | Distributed Energy Resource | |
| DG | Distributed Generation | |
| DM | Dynamic Moderation | |
| DR | Dynamic Regulation | |
| DSF | Demand-Side Flexibility | |
| DSO | Distribution System Operator | |
| EBGL | Electricity Balancing Guidelines | |
| EDF | Explicit Demand Flexibility | |
| EMD | Electricity Market Directive | |
| ENTSO-E | European Network of Transmission System Operators (union of TSOs in Europe) | |
| ERPS | Enhanced Reactive Power Service | |
| EU | European Union | |
| FCR | Frequency Containment Reserve | |
| FFR | Firm Frequency Response | |
| FSP | Flexibility Service Provider | |
| GME | Gestore dei Mercati Energetici (NEMO in Italy) | |
| GSE | Gestore Servizi Energetici | |
| HEP-ODS | Hrvatska Elektroprivreda - Operator Distribucijskog Sustava (DSO in Croatia) | |
| HOPS | Hrvatski Operator Prijenosnog Sustava (TSO in Croatia) | |
| HORTE | Hrvatski Operator Tržišta Energije (NEMO in Croatia) | |
| ID | Intraday market | |
| IDA | Intraday Auction | |
| IDAE | Instituto para la Diversificación y el Ahorro de la Energía | |
| IDF | Implicit Demand Flexibility | |
| kW | Kilo Watt | |
| kWh | Kilo Watt hour | |
| LV | Low Voltage | |
| MFR | Mandatory Frequency Response | |
| mFRR | Manual Frequency Restoration Reserve | |
| MSD | Mercato del Servizio di Dispacciamento (Ancillary services market in Italy) | |
| MV | Medium Voltage | |

| | |
|---|---|
| MVar | Megavolt Ampere Reactive |
| MVArh | Megavolt Ampere Reactive hour |
| MW | Mega Watt |
| MWh | Mega Watt hour |
| NEMO | Nominated Electricity Market Operators |
| NGESO | National Grid Electricity System Operator (TSO in the UK) |
| OMIE | Operador del Mercado Iberico – Polo España (NEMO in Spain and Portgual) |
| ORPS | Obligatory Reactive Power Service |
| REE | Red Eléctrica de España (TSO in Spain) |
| RR | Replacement Reserve |
| SDAC | Single Day-Ahead Coupling |
| SIDC | Single Intraday Coupling |
| SO | System Operator |
| STOR | Short-term Operating Reserve |
| TOU | Time-Of-Use tariff |
| TSO | Transmission System Operator |
| UK | United Kingdom |
| USEF | Universal Smart Energy Framework |
| UVAM | Virtually Aggregated Mixed Units |
| VLP | Virtual Lead Party |

# 1. INTRODUCTION

Transitioning to a more sustainable energy system requires adopting both renewable energy technologies and distributed generation (DG). Renewable energy technologies, such as wind and solar power, provide clean and abundant sources of electricity, while distributed generation allows for the generation of electricity at or near the point of use, reducing transmission losses and increasing reliability.

However, the main drawback of these resources is that their energy production depends on an energy source that is not controllable. An electricity demand change could cause imbalances and constraints in the transmission and distribution grid. In [1], it is simulated different distribution grids in two locations in Europe with high penetration of DG, and it is determined that grid contingencies are very likely to occur, especially related to voltage problems.

New manageable resources are appearing on the consumers' side, called Distributed energy resources (DERs). These resources gather DG and demand-side electrification (e.g., electric vehicles, energy storage and other manageable consumptions [2]). The use of DERs changes the position of consumers from passive actors to active elements of the electricity system, also known as a prosumer, giving energy independence to the consumers and, at the same time, changing its demand profiles.

To maximise the energy produced by variable renewable generators, it is necessary to match energy generated by renewable generators and demand, if not frequency problems or other technical problems (as the ones described before) could occur. The correct use of energy resources provides the necessary system flexibility to manage imbalances and respond quickly to variations in supply and demand [3]. This leads to a reduction in global costs, including resolving congestion in the grid. As defined in [4], "a flexible grid is the one that can respond reliably and quickly to large fluctuations in supply and demand".

In the past, flexibility was provided by thermal generation, cross-border interconnections and pumped storage hydropower [3]. With the new paradigm in which renewable generators will lead the generation schemes, in the case of requiring more generation due to an unbalance, these technologies could not fully provide it. It is here when the concept of Demand-Side Flexibility (DSF) appears. DSF is the ability of DERs to shift or change their expected consumption pattern in response to a signal [5].

DSF can be divided into implicit and explicit demand flexibility (IDF and EDF, respectively). IDF incentivises consumers to change their energy consumption patterns through dynamic tariffs, such as time-of-use (TOU) tariffs, by shifting customers' consumption in low-cost periods due to, for example, lower utilisation of the distribution network or due to high penetration of renewables. On the other hand, there is EDF. In this type of DSF, the modification of customers' consumption is incentivised not because of electricity tariffs but because customers receive direct economic benefits (savings on electricity bills or even incomes) by shifting their consumption in a given period. More information about the difference between both concepts can be found in [3].

The role of the aggregator appears to make possible the inclusion of prosumers into different markets [3], [6]. Aggregators gather energy resources from different prosumers (mainly demand) and bring the flexibility of these resources to different electricity markets. Then, aggregators trade this available flexibility with different system agents (DSOs, TSOs or retailers) to ensure the system's correct operation or its position in the day-ahead and intraday markets to obtain more profits. An aggregator could be an energy retailer or a third-party company, known as Independent Aggregator. Participation of demand flexibility and aggregation is encouraged by European directives (Directive 2019/944 [7]) and policies like the Clean Energy Package [8].

Some papers in the literature have analysed how the markets are, but usually only one type of market is analysed, like in [9] which analyses balancing markets in Europe or in [10], which analyses ancillary markets of five countries, but other markets and participation of demand is not clearly described.

There are existing studies about one specific country analysis, like in [11], or about how it is structured the electricity system as in [12], [13] or [14], but it is no available information about demand participation in those markets. Papers about demand participation in electricity markets, as in [15], only describe balancing markets, but the rest markets in which there is potential participation of consumers are not described.

This study focuses on four countries: Spain, Italy, Croatia and the United Kingdom. These four countries have different development of electricity markets among them but are similar to other countries around them. For instance, Spain shares many characteristics with the Portuguese system. Regarding the development status of each country, e.g. the UK has followed a long development process of its electricity market, reaching up to three reforms of its system [14], while on the other side is Croatia, where the liberalisation of the market started in 2008.

The goal of this paper is to review the different electricity markets in the countries proposed, first by analysing their regulation and then considering the participation of prosumers in the different markets. In order to compare the openness of each market between different countries, a methodology is proposed using a scale of 0 to 3 for each market and country.

This paper describes all potential markets where prosumers can participate, as defined in section 2. After that, the current situation of prosumers' participation in each of the selected countries and their barriers are described, indicating if it is possible that prosumers can participate in each different market and if it is possible to participate in a collective self-consumption scheme. Then, a methodology to analyse the openness grade of participation of prosumers is proposed in section 3, looking at the different aspects found in the previous analysis. Section 4 applies the methodology previously defined to the four countries reviewed.

## 2. EUROPEAN ELECTRICITY MARKETS REVIEW AND OPENNESS ANALYSIS

This section describes all potential markets in which prosumers can participate: day-ahead (DA) and intraday (ID) markets, ancillary services, adequacy services and constraint management markets and the local flexibility market. It is also analysed the possibility of using collective self-consumption schemes. This is not a market but a way to make the prosumer a participant in the electricity system, so it will also be studied.

A description of the functioning of each market is presented to explain whether and how participation is possible in each market. In all markets analysed, it is explained how these markets work, who is responsible for them and how prosumers can participate in them. In Local flexibility markets and collective self-consumption sections, the participation of prosumers is implicit in these markets, so no subsection on openness to prosumers will be made.

### 2.1. Day-ahead and Intraday markets

The USEF framework defines that explicit flexibility could be used in the wholesale markets in different ways. For example, Balancing Responsible Parties (BRPs) can use explicit flexibility to optimise their portfolio, e.g. when BRPs detect an error in their forecast and needs to adjust their position in the market, they can adjust their position by using this flexibility, avoiding the corresponding imbalance charges [5].

European wholesale markets are very similar to each other, mainly because all are under the Single Day-Ahead and Intraday Coupling (SDAC and SIDC, respectively). These platforms were to make possible a single pan-European electricity market, using a standard price coupling algorithm (PCR EUPHEMIA) to calculate electricity prices in the different national markets, considering the cross-border transmission capacities [16]. The common characteristics of the day-ahead, continuous and intraday auction markets in Europe are described in Table 1.

*Table 1 Summary of common characteristics of European day-ahead and intraday markets*

| Terminology | | Day ahead market | Intraday Continuous Market | Intraday Auctions Market |
|---|---|---|---|---|
| **Pricing Method** | | Marginal Pricing | For a buy offer, if the price is higher than any of the sell offers, the bid is accepted.<br>For a sell offer, if the price is less than any of the buy offers, the bid is accepted.<br>*The price is set by the purchase offer when the matching is made.* | Marginal Pricing |
| **Algorithm** | | EUPHEMIA | LTS and SIDC Solution | At this moment, each country has its algorithm. SIDC solution will be implemented in the future [17]. |
| **Market frequency** | | Daily auctions | Hourly, Half-hourly and Quarter-hourly (according to the bidding zone) | According to the bidding zone. |
| **Bidding period** | | No opening gate and closing time at noon CET on the day before power delivery (D-1) | Gate opening time: 15:00 a.m. D-1, and gate closure time: h- 60 min/ 30 min/ 5 min according to bidding zone | According to the bidding zone. |
| **Product characteristics** | *Granularity* | Hourly/30-min (according to bidding zone). | Hourly/30-min/15-min (according to bidding zone). | Hourly/30-min/15-min (according to bidding zone). |
| | *Minimum and maximum bid price:* | - 500 €/MWh and 4.000 €/MWh | -9.999 €/MWh and 9.999 €/MWh. | |
| | *Minimum and Maximum Quantity to bid* | 0,1 MW and No constraint | | |
| | *Price Tick* | 0,1 €/MW | | |

### 2.1.1. Spain

Day-ahead and Intraday in Spain are organised by OMIE (*Operador de Mercado Iberico – Polo Español*), the Iberian Nominated Energy Market Operator (NEMO). Day-ahead and intraday markets are where most of the energy is procured. It is acquired more than 70% of the energy consumed in Spain in 2020 [18].

The intraday auction market is divided into six bidding sessions (IDA 1-IDA 6) starting at 14:00 the day before delivery [19]. These sessions are opened once the Spanish TSO, *Red Eléctrica de España* (REE), publish the Final Viable Daily Program.

At present, OMIE's own matchmaker is used for the day-ahead and intraday market, but in the future, with the modification of the intraday auction market, the Euphemia algorithm will be used.

Intraday continuous market in Spain has been connected to the SIDC since 2018, with the first go-live wave. There is a total of 24 bidding rounds, one for each hour of the day, to trade energy for the current and the following day, and each round is closed one hour before the hour of delivery.

#### 2.1.1.1 Market openness to prosumers

There is no evidence of explicit flexibility participation of small prosumers in wholesale markets. However, electro-intensive industries can participate in the market directly or through a representative.

Regarding small prosumers, they can participate in markets using the implicit flexibility of their appliances in the different electricity tariffs offered.

Spanish retailers offer modalities of tariffs: fixed price with up to three time periods, indexed price to day-ahead market and flat rate tariffs. Furthermore, in Spain exist a regulated tariff, known as PVPC. This tariff reflects the prices of the spot market in the electricity cost (plus other expenses), and five electricity retailers offer it. About 35.1% of Spanish domestic consumers have contracted this tariff [20]. A consultation is open to determine how to form the price of these tariffs to make them more stable [21].

According to Spanish Electricity Sector Law 24/2013, energy retailers can offer dynamic price contracts [22]. Electric tariffs in Spain include a dynamic cost of the usage of distribution and transmission grids (known as tolls) and other charges like the extra cost of production activity in electricity systems in non-mainland territories, among others. This dynamic cost is divided into three periods, with higher costs during peak consumption periods, as established in Royal Decree 148/2021[23].

Regarding the revenues generated by self-consumption surplus, prosumers can get compensated for this surplus of energy injected for energy consumed in other periods. The energy purchased from the grid will be paid with the surplus energy injected at the average price of the hourly market minus the cost of deviations (for PVPC tariff) or a price agreed with the energy retailer [24].

This method is not considered as a net metering schema because the surplus of energy consumed during hours without injection has to pay the transmission and distribution costs, the prosumers only save the energy costs.

For the case of more significant consumers, energy procurement is usually done by using real-timing tariffs, indexed to the price of the wholesale market, or as it has been sentenced before, by participating directly in these markets.

### 2.1.2. Italy

The Italian NEMO, GME (*Gestore dei Mercati Energetici*), regulates the day-ahead market. Most of the energy consumed in Italy is traded in this market, and, unlike in Spain, Italian electricity is traded for each geographical zone (North, Central North, Central South, South, Calabria, Sicilia and Sardegna).

Regarding the intraday markets, the intraday auction market is divided into three sessions: MI-A1 closes at 15:00 D-1, MI-A2 closes at 22:30 D-1, and MI-A3 closes at 10:30 D. Meanwhile, the continuous intraday market, known as MI-XBID, is divided into three phases: Phase I opens at 3.30 p.m. the day before the delivery (D-1), Phase II at 10.30 p.m. D-1 and Phase III at 10.30 a.m. D.

#### 2.1.2.1 Market openness to prosumers

In Italy, Law 124/2017 introduces the obligation on retail suppliers to provide at least one commercial offer linked to wholesale spot market prices [25], known as PLACET tariff. Some suppliers give specific contracts with dynamic hourly prices for small customers with a smart meter installed.

As in Spain, ARERA (Italian regulator) offers a regulated tariff called *Servizio di Maggior Tutela* (Higher Protection Service). *Servizio di magior tutela* is an option provided in the Italian energy market, which guarantees the consumer the supply of electricity and gas at the economic and contractual conditions established by the ARERA. This tariff is divided into two price slots, F1 from 8 a.m. to 7 p.m. and F23 from 7 p.m. to 7 a.m., being the last the cheapest. Around 33% of the consumers in Italy buy their energy in the *Servizio di Maggior Tutela* [26].

Regarding remuneration for the injection of energy surplus of prosumers, there are two possibilities: *Scambio sul Posto* or *Ritiro Dedicato.*

The first case is a net metering mechanism, *Scambio sul Posto*. Net metering allows prosumers to inject the surplus electricity produced by an electricity production plant (up to 200 kW) into the grid and then consume it later to meet prosumers' electricity consumption. Then, GSE (*Gestore dei servizi energetici*) will pay the prosumer for energy consumed in hours without injection at the price determined in the electric tariff [27].

In the case of *Ritiro Dedicato*, it allows the user to sell all surplus to the GSE at a determined price, receiving these incomes from the prosumers and not being discounted from the energy bill [28].

No information about the direct participation of small prosumers in the day-ahead and intraday market in Italy has been found.

### 2.1.3. Croatia

Croatian NEMO, HROTE, organises the Day-Ahead Market in Croatia. In this market, members submit orders to buy or sell electricity for 24 hours of the following day. This market is coupled with the European SDAC market via the Croatian-Slovenian border.

The unique Intraday market in Croatia is a continuous intraday market. This market was created on the second go-live wave launched for SIDC on November 2019.

### 2.1.3.1 Market openness to prosumers

Croatia's retail market is less developed of the four analysed: it only offered flat rate tariffs with one or two prices depending on whether the consumer has a smart meter installed.

Since 2008, Croatian customers can choose the energy retailer freely. Currently, there are seven retailers in Croatia. All of them offer electricity tariffs in the liberalised market except for HEP – Opskrba, which is the one that provides a regulated tariff for households (known as Universal service). 88% of the energy consumed by households is supplied through public service [29].

In the case of self-consumption surplus, energy suppliers must purchase the excess electricity injected into the grid by prosumers at the 90% price of the supply tariff (net-billing) [30].

There is no information available about the involvement of small prosumers in the day-ahead and intraday market in Croatia.

## 2.1.4. United Kingdom

Two different Market Operator operates the day-ahead and intraday markets, these are EPEX and Nord Pool. In those NEMOs, day-ahead and intraday services are traded. Regarding the day-ahead market, products sold can be hourly and half-hourly. Apart from this, the transmission capacity of the interconnection of Great Britain with continental Europe (Netherlands, France and Belgium) is auctioned in both marketplaces.

The intraday market is structured in intraday auctions and continuous markets. The intraday auctions market is divided into two sessions: IDA1 and IDA2. The continuous intraday market is based on the constant trading of products with up to half-hour granularity and is organised by Nord pool and EPEX Spot.

### 2.1.4.1. Market openness to prosumers

British electricity suppliers offer different types of tariffs. These are Standard variable tariffs (prices in these tariffs can vary to reflect changes in the wholesale prices), fixed-rate tariffs, Economy 7 (two different prices, an on-peak rate and an off-peak for seven hours of the night) and indexed tariffs.

Standard variable tariffs are a specific type of tariff in which Ofgem (British regulator) defines a price cap that electricity suppliers are not allowed to exceed, which could be considered a regulated tariff. The prices of this type of tariff can be lower but not higher than this cap. This cap is defined every six months, and some energy retailers must offer it.

According to Ofgem data, 60% of domestic electricity customers have contracted a standard variable tariff, 38% have a fixed tariff, and the 2% remaining have other tariffs [31].

For the remuneration of the surplus of prosumers with self-consumption, the British regulator Ofgem has defined a new scheme that substitutes the previous feed-in-tariff service, this new initiative is called Smart Export Guarantee. This initiative obligates energy retailers to pay surplus energy of prosumers on the conditions defined by energy retailers [32].

Direct participation of small customers in the UK was not found, but the Balancing and Settlement Code modifications are in process to allow aggregators' participation in the British wholesale market [33].

### 2.2. Ancillary services, adequacy services and constraint management markets

Another of the markets with the potential to use explicit demand flexibility, according to USEF, are ancillary, adequacy, and constraint management services markets [5].

Ancillary services are defined in the Electricity Market Directive (EMD) as the services transmission and distribution systems required to operate appropriately [7]. These services include balancing and non-frequency ancillary services.

Balancing services are essential for maintaining the grid's frequency within limits established by the TSO to avoid the risk of a global blackout. ENTSO-E defines a stability range for the frequency between 47.5 and 51.5 Hz [34]. At European level, there are defined specifics services for balancing the electricity system, these are described in Table 2.

*Table 2 Definition of balancing products at European level [35] and [36].*

| FCR – Frequency Containment Reserve | FCR instantaneously balances out frequency deviations and is activated automatically. Assets that provide FCR need to react within 30 seconds to the steering signals fully. |
|---|---|
| aFRR – automatic Frequency Restoration Reserve | aFRR is automatically activated after FCR, replacing the assets triggered by FCR. aFRR has to be fully activated within 5 minutes. |
| mFRR – manual Frequency Restoration Reserve | mFRR is activated to substitute assets involved in aFRR. This service is manually activated and must deploy all its capacity in less than 15 minutes. |
| RR – Restoration Reserve | The reserve replacement process replaces the activated mFRR.<br>Activation time varies between different European countries (30 minutes to hours). |

The Electricity Balancing Guidelines (EBGL) target the establishment of common European platforms and, as a result, the standardisation of European balancing market processes [35]. For each balancing service, the EBGL requires the definition of European platform for trading normalised balancing products around Europe. These platforms are: Platform for the International Coordination of Automated Frequency Restoration and Stable System Operation (PICASSO) – for aFRR service; Manually Activated Reserves Initiative (MARI) – for mFRR service and Trans-European Restoration Reserves Exchange (TERRE) – for RR service.

Non-frequency ancillary services are used to control other parameters necessary to maintain the quality of the electricity provided through the transmission and distribution grid. The services within this category are steady-state voltage control, fast reactive current injections, inertia for local grid stability, short-circuit current, black start capability and island operation capability. The name and number of services may change depending on the needs of the country under analysis.

Adequacy services, mainly capacity markets, are used to ensure enough energy generation in the future. Capacity markets are a tool used to procure long-term capacity contracts to ensure energy security, with power plants receiving capacity payments in addition to revenue generated from the sale of electricity on the energy market. These markets are present in only 11 European Union countries [37].

Lastly are the constraint management services. Grid operators use these services to maintain the operation of the grids without physical constraints. Different services are used according to problem solve. The primary services required in this topic are voltage control and congestion management, e.g. grid saturation due to excess demand in one grid node or voltage problems.

### 2.2.1. Spain

Ancillary services in Spain are composed of balancing services, solutions to technical constraints (constraint management), voltage services and capacity markets. The first service, balancing services formed by secondary regulation (aFRR) , tertiary regulation (mFRR) and Replacement Reserve (RR), is based on a market basis; meanwhile, primary regulation (FCR) is a mandatory service that can be only provided by spinning generators (e.g. thermal generators and hydropower) [38], [39].

In Replacement Reserves and tertiary regulation, energy mobilised (€/MWh) is remunerated based on a pay-as-clear scheme. Regarding secondary regulation, capacity, also known as capacity band or regulation band, is paid in a marginal market (€/MW), while energy is remunerated at a price cleared in the tertiary regulation [40]. Balancing services are activated only if deemed necessary following negotiations in the wholesale markets, and multiple services may be activated as needed.

Spain has two constraint management services: Solution of technical constraints and Supplementary voltage monitoring service for the transmission network. Solution of technical constraints is used to solve any circumstance or incidence derived from the production-transport system situation that affects the electricity system's safety, quality and reliability conditions. This service is divided into two parts: one after the publication of the Basic daily programme of operation and one in real-time [41]. Remuneration in the solution of technical constraints, the system operator (REE) will conduct an economic assessment of the possible solutions and choose the offers that represent the lowest cost [40].

Regarding the voltage service, it is mandatory that all thermal and hydropower generators with a power higher than 30 MW, consumers with a contracted power higher than 15 MW and DSO and TSO have to provide this service. Each month, all participants in this service will be paid for the capacity and energy offered to absorb or generate reactive power [42].

The capacity market was proposed in 2020 by the Spanish Government to incentivise investments in different technologies that can provide stability to the electricity system by establishing a framework in which price signals are set in the different auctions done [43].

These services will be contracted through competitive bidding procedures managed by the operator of the "pay-as-bid" auction system, ensuring that this market is technologically neutral, allowing generation, storage, consumers and self-consumption to participate in this market. Two modalities of auctions with different time frames have been proposed, one year of service provision and between one and five years of provision. At the moment of writing this paper, no public information has been found regarding the status of this capacity market proposed.

### 2.2.1.1. Market openness to prosumers

From 26th January 2021, all market participants with generation, demand or storage with a minimum supply capacity of 1 MW can be Balancing Service Providers (BSPs) [44].

Each BSP has to fulfil the information exchange requirements with the system operator: structural information and telemetry in real-time.

Different control and specific tests for each balancing service: secondary has more challenging requirements than tertiary and RR (technical requirements are the same for the last two services, both for generation and demand). In addition, to take part in the aFRR market, at least 200 MW are needed to form a regulation zone, which is necessary to take part in aFRR services, hindering the participation of aggregators due to the high threshold.

Aggregation is only allowed through an energy retailer, independent aggregation is not yet allowed, and it is expected to be adopted this role in Q1 2023 [45].

According to [46], only one unit has taken part in the ancillary markets in Spain, and it was for only one hour in the RR service.

A new service called *Servicio de respuesta activa de la demanda* (Active demand response service) has been published to be used when there is a lack of manual balancing reserves (mFRR and RR). Providers of this service will reduce their energy consumption by up to three hours [47]. Only could take part in this service consumers with at least 1 MW of demand, excluding small consumers from this service. 497 MW was procured in the first auction with a price of 69.97 €/MWh [48].

Regarding the capacity market and constraint management service, no information on the possibility of prosumers' participation in these markets has been found.

### 2.2.2. Italy

Terna, the Italian Transmission System Operator, purchases resources from the Ancillary Services Market (MSD) to manage and monitor the system, relieve intra-zonal congestion, create an energy reserve, and perform real-time balancing. There are two markets within this service: ex-ante MSD and Balancing Market.

There are six scheduling sub-phases in the ex-ante MSD. Terna accepts offers to buy and sell energy on the ex-ante MDS to relieve remaining congestion and generate reserves (so-called re-dispatching action). The Balancing Market allows balancing providers to submit offers continuously until one hour before delivery day.

Both markets trade frequency regulation (aFRR, mFRR and RR) and congestion resolution resources (known as *Risorse per la risoluzione delle congestioni in fase di programmazione*). Only generation units connected to the transmission grids with a power higher than 10MVA can participate in ancillary services. Terna compensates for all offers accepted in the services described before the price offered (pay-as-bid) [49].

Apart from these markets, Terna also uses other ancillary services to control grid stability and restore emergency scenarios. These services, such as primary reserves, voltage regulation, emergency, and restoration, are mandatory for some agents in the electricity system, mainly large thermal generators.

Regarding capacity markets, Terna uses this market to procure long-term contracts awarded through competitive bidding. Participation is voluntary and technology-neutral, so demand and generation can participate in this market. The operators chosen must offer their capacity to energy and services markets, are remunerated with an annual fixed premium and have to repay (if it is positive) between the price on the participant markets and the price defined by ARERA.

### 2.2.2.1. Market openness to prosumers

To facilitate the participation of new resources in the ancillary services market (specifically mFRR), in 2017, Terna started a pilot project known as UVAM (Virtually Aggregated Mixed Units) [50].

UVAM allow participation in the tertiary reserve service (mFRR), not only generation but also demand and energy storage systems, mainly distributed resources.

Participant units have to have a capacity of at least 1 MW. Remuneration is based on a pay-as-bid scheme linked to the energy activated (€/MWh) and availability (€/MW).

Information in [51] indicates that only a few demand units have taken part in UVAM and are mainly industrial consumers (mixed with traditional generation).

For the capacity market, prosumers can participate in an aggregated way. However, de-ratings factors inhibit participation. In the last auction, 30% of new capacity was awarded to non-traditional providers, largely front-of-meter batteries [52].

### 2.2.3. Croatia

Croatian TSO and DSO (HOPS and HEP-ODS) are responsible for procuring ancillary services, which are: automatic frequency restoration reserve (aFRR), manual frequency restoration reserve (mFRR) and voltage power regulation [53].

All these services are remunerated to the units that are activated. Procurement is done periodically through a public tender (monthly, weekly, daily and/or intraday).

Regarding balancing products, primary frequency regulation (FCR) must be provided by generators, while the rest (aFRR and mFRR) are contracted to balancing providers by the system operator.

Remuneration in aFRR and mFRR is based on two concepts: capacity (MW) defined in a marginal tender and energy used (MWh) compensated at the price specified by the energy in mFRR [54].

Providers taking part in the balancing services should have a power higher than 1 MW and have a common information centre (in the case of more than one unit participating) that coordinates individual units and is communicated with HOPS [55].

In voltage regulation service producers, end customers, transmission and distribution networks participate in maintaining voltage stability. The choice of reactive power supplier is based on technical requirements, minimum costs and ensuring the availability of reactive power reserves in certain parts of the transmission network. All users from the grid that can supply reactive power and are interested in providing this service need to have an agreement with the transmission system operator.

At the time of writing this paper, no capacity market or intention to create one has been identified.

In case of the system operator makes a forecast of congestion on a transmission network, the measures considered in the Croatian transmission regulation are: implements a change in the switching state of the transmission network, requires the amendment of electricity production plans, limits the consumption of network users or limits the allocation and use of cross-border transmission capacities.

### 2.2.3.1. Market openness to prosumers

The figure of the aggregator and independent aggregator are present in the Croatian electricity market lawspecifying that any type of BSP (aggregation, generation, demand and storage included) can participate in ancillary services [56]. There is already two companies registered as aggregators in Croatia: KOER and Nano Energies. KOER has already been accepted to provide flexibility services in mFRR market using aggregated demand from industries [57].

No information about the participation of prosumers in constraint management services has been found.

### 2.2.4. United Kingdom

Ancillary services requested by British System Operator, National Grid ESO (NGESO), are divided into different service categories: Frequency response service, Balancing Mechanism, Reserve services, system security services and reactive power service.

Frequency response services are used in the British grid to control the real-time balance between generation and demand. It is equivalent to the FCR service in the European Union. The products available in this service are: Dynamic containment (DC), Dynamic Moderation (DM) and Dynamic Regulation (DR); Firm Frequency Response (FFR); and Mandatory Frequency Response (MFR).

The main difference between frequency response products is the acquisition time, i.e. FFR is acquired a month ahead, while DC is tendered a day ahead [58].

In addition, there are three different response times for frequency response services, these are Primary response (activated 10 seconds after an event, sustained 20 seconds),

Secondary response (activated 30 seconds after, sustained up to 30 min) and High frequency response (10 seconds after the effect and can be sustained indefinitely) [59].

DC was launched in 2020 as a service-remunerated pay-as-bid, but in 2022 has changed to pay-as-clear. DM and DR have been launched in 2022. It has to be emphasised that these services have different auctions depending if the frequency is higher or lower than 50 Hz, i.e. the service auctioned to serve reserves when the frequency is higher than 50 Hz, then is known as High (e.g. DM High Frequency or DM HF). These services are described in Table 3.

*Table 3 Description of characteristics of DC, DM and DR [60].*

| Services | DC | DM | DR |
| --- | --- | --- | --- |
| Product Characteristics | Post-fault service to contain frequency between +/-0.5Hz range in the event of a sudden generation loss or demand. It only occurs when the frequency moves outside operational limits (+/-0.2Hz). | Designed to assist frequency management following large imbalances. The aim is to contain frequency within operational limits of +/-0.2Hz. | Designed to correct minor continuous deviations in frequency slowly. The aim is to regulate frequency around the target of 50Hz continually. |

Another of the frequency response service is Firm Frequency Response (FFR). FFR is divided into two types: non-dynamic and dynamic. The non-dynamic service is triggered at a determined frequency deviation, and a response is not necessary if this range is not exceeded. Tenders in FFR are done monthly and are paid an availability fee (£/h), a response energy fee (£/MWh), and even some units state a nomination fee (£/h). Price is defined on a pay-as-bid basis. Balancing Market and non-Balancing Market units can provide these services, being 1 MW the minimum bid size [61].

Mandatory Frequency Response (MFR) is a service triggered automatically in large generators in response to a frequency change. It is obliged to provide this service to generators located in the National Grid with a power higher than 50 MW, 30 MW or 10 MW, depending on the generator's location. A holding (£/h) and response energy (£/MWh) payment is given to all unit mobilised [62].

Another way that the system operator can balance the transmission system is by using the Balancing Mechanism (BM). In BM, balancing mechanism units and non-balancing mechanism units send bids (increase generation or reduce demand) and offers (reduce generation or increase demand) after the Gate Closure to provide reserves to the grid. NGESO will choose the bids that better satisfy the balancing requirements, considering the power stations and transmission system's technical limitations [63]. Units suitable for this service are selected in merit order and paid as bid (£/MWh).

Reserve services are used to balance supply and demand on short timescales in case of problems with some generators or unexpected increases in demand. The leading reserve service in the UK is Short-term Operating Reserve (STOR). The service is open to any technology that can increase generation or reduce demand by at least 3 MW. Providers must respond to an instruction within 240 minutes, although response times within 20 minutes are preferred. As in FFR, BM and non-BM can participate in this market.

STOR reserves [64] are procured in the day-ahead market through a daily pay-as-clear auction process. The auction closes at 05:00 for service delivery the following day at 05:00. Providers submit their availability prices and MW offering before the day-ahead auction, which is then cleared to secure capacity for STOR at the lowest availability cost. A STOR bid is accepted when NGESO considers that the total costs of securing and operating the system are lower with the bid than without it.

BM and STOR will be equivalent to European mFRR and RR services according to the response time of these services.

Restoration service is used in the case of a total or partial shutdown of the British transmission grid. This service can be procured from Power Generation Modules, HVDC systems and distributed Energy Resources that can restart from shutdown without reliance on an external supplier. Payments are based on the provider's availability (£/settlement period).

There are two reactive power services, the Obligatory reactive power service (ORPS) and the Enhanced Reactive Power Service (ERPS). In ORPS, any generator can be requested to produce or absorb reactive power to manage system voltage to the desired value. Remuneration is based on a utilisation fee (£/MVArh)[65]. The other reactive service is ERPS, in which any installation that can offer this service is allowed to take part. In this case, generators can get incomes from available capability price (£/MVar/hr), synchronised capability price (£/MVar/hour) and/or a utilisation price (£/MVArh) [66].

Regarding capacity markets in the UK, NGESO procures capacity (£/kW/year) ahead of delivery to ensure sufficient investment in the development of new generation to meet ongoing reliability standards. Participants are remunerated based on a price set at the auction clearing price. Technology-specific de-rating factors apply to the capacity secured in each auction which de-rates the MW on which remuneration is based according to the potential contribution of the selected technology to energy system security at times of stress [67].

There are two capacity auctions each year: the T-4 auction (buys capacity needed for delivery in 4 years and secures agreements for up to 15-year terms) and the T-1 auction (used to top-up capacity and ensure there is sufficient generation and demand-side response to balance the electricity system each year).

Lastly, constraint management in the UK is carried out through bilateral agreements and through the BM, being the most used the last option [68]. Regarding the bilateral agreements, these will be published by NGESO once a possible congestion in the transmission grid is identified, being the system operator who contacts potential providers [69].

### 2.2.4.1. Market openness to prosumers

Aggregators and independent aggregators have been allowed in the British electricity market since the role of Virtual Lead Parties (VLPs) was introduced. This agent is created to enable the participation of aggregated resources (independently from a retailer) in the Balancing Mechanism [70].

In FRR and STOR, conditions make possible the participation of prosumers by aggregating their demand into VLPs if bids are higher than 1 and 3 MW, respectively.

In 2020 energy supplier Social Energy won the first domestic FFR contract from a portfolio of hundreds of households with batteries. However, some FFR participation requirements remain onerous or unclear for domestic portfolios (e.g. testing regime) [71].

In early 2021 Ofgem approved Balancing and Settlement Code Modification P375, which allows individual asset meters to be used for settlement purposes, opening the BM for smaller participants [72].

At the end of 2022, NGESO launched a test service named Demand Flexibility Service in which consumers with a smart meter can take part by reducing their consumption for one to four hours, with prior notification of their suppliers, getting revenue up to 3 £/kWh [73].

Lastly, in capacity markets auctions, DSF and aggregator participation is permitted, but in practice, substantial bureaucratic challenges often deter smaller assets and portfolios, which cannot justify the effort of the involvement.

### 2.3. Local flexibility markets

As presented in the introduction, local flexibility markets are other potential markets in which prosumers can participate.

The development of flexibility markets is defined in the Clean Energy Package and regulated in the EMD, which states that Member States have to change their regulation to allow and incentivise DSOs to make use of flexibility services (including constraint management), provided by DERs, in order to improve the efficient operation of their networks [7].

The regulatory framework should establish a system that allows distribution system operators to acquire services such as distributed generation, demand-side flexibility, and energy storage from providers. Additionally, the framework should encourage the implementation of energy efficiency measures that can cost-effectively reduce the need for upgrades or replacements of electricity capacity and assist in the secure and efficient operation of the distribution system.

These services, remarks EMD, have to be purchased as constraint management services based on market-based procedures. Being these services purchased in the so-called Local Flexibility markets.

#### 2.3.1. Spain

OMIE and IDAE (*Instituto para la Diversificación y el Ahorro de la Energía*) have been developing since 2019 a local flexibility market project called IREMEL [74]. The IREMEL project aims to create a local flexibility market to incentivise DSOs to purchase flexibility services from DERs (distributed renewable generation, demand and energy storage) to avoid future congestion problems. This market is currently in a testing period for its upcoming launch. To solve these problems in the distribution grid, two solutions are envisaged in the IREMEL project:

- Short-term market: products are traded on a day-ahead and intraday basis.

- Long-term market: procurement of products is made months in advance.

Short-Term markets are organised in two modes: free (continuous market) and Short-term Local Market on request of the DSO.

The short-term free market is conceived as a market in which DERs can participate in a continuous market to trade energy within the same zone. This market is a continuous intraday market but is planned to be used in isolated systems, i.e. an island.

The other mode of short-term markets is the short-term market on request of the DSO. This mode is created to solve specific network congestions arising from peak demand or generation in an area or to respond rapidly to unforeseen situations or planned short-term actions.

Within this mode, two different products are traded: daily and intraday. These products are traded in an auction the day before activation or on the same day, and are paid only for the amount of energy activated (€/MWh). The allocation of intraday products can be as much as 15 minutes before the activation of the service.

On the other hand, it is the long-term market. This market resolves congestion or permanent problems and delays investments in network extensions. It also provides an availability payment in this market, which will help new distributed resources emerge and create new investments from prosumers.

In this market, two products are being tested: Fixed activation and availability products. These products proposal is fully configurable to the needs of the customer. In both products, DERs are awarded with a remuneration for their availability (€/MW) and activation (€/MWh). Fixed activation product defines an agreement between the DSO and DERs to provide a determined service for a specified period and price. Regarding the availability product, a price is defined for availability and activation. Then, activation price is used in the short-term market to compete with the rest of the offers and, if the bids from the long term service are cleared or there are no other bids than the ones from the long term, the awarded long term DER will provide this service.

To participate in this market, DERs (potentially within an aggregator) have to demonstrate that they have the financial resources to take part in the market (known as *Garantías*), that their resources are located within the traded area and their technical capabilities. All these conditions must be qualified by the market operator (OMIE) and the relevant distribution system operator (DSO) where flexibility services are required.

The market operator is who activates the negotiation period for the services required by the different DSOs, being them the responsible for communicating the different requirements that may be needed, both in the long and short term, either as upward or downward requirements, as well as being in charge of communicating to the market operator whether the different participating DERs have fulfilled the past requirements.

### 2.3.2. United Kingdom

Energy Networks Association, the representative of owners and operators of transmission and distribution networks, presented in 2017 the Open Network Project in which was defined a range of standardised active power services to be used by DSOs to reduce costs to customers [75]. These four products are:

- Secure service is used to control network peak demand loading and minimise network load in advance;
- Dynamic service has been developed to support the network in specific fault conditions, such as during maintenance work;
- Restore is designed to assist with network restoration in a rare network failure alleviating network stress;
- Sustain is a scheduled constrain service that manages peak demand loading on the network and pre-emptively reduces network loading.

These products could be used by DSOs in order to solve different problems in their networks. Two marketplaces are available to trade these products: Flexible Power and Piclo Flex, these are similar to the IREMEL project in Spain.

Flexible power is an initiative from five British Distribution Network Operators (Western Power Distribution, Northern Powergrid, Scottish and Southern Electricity Networks, SP Energy Networks, and Electricity North West) to procure flexibility from customers in order to avoid constraints in distribution networks using assets at all voltage levels [76]. Flexible power uses a customer-facing portal where Flexibility Service Providers (FSPs) can declare their flexible assets' availability, receive dispatch signals, and view performance and settlement reports.

The flexibility services defined above were the only ones available in Flexible Power, but new long-term services have been opened to bring new opportunities for different assets to enter. These new services are Secure long-term and Dynamic long-term. Short-term products negotiated in Flexible Power are procured one week in advance; meanwhile, long term is procured months ahead.

With the addition of the three new products, six products are available in the Flexible Power market. It has been decided to allocate zone as Secure or Dynamic zones to simplify the number of products available in each area. Sustain and Restore would be available in both zones. These six products are remunerated differently: Sustain and Restore are paid at a fixed price, while secure and dynamic are paid-as-clear.

On the other side is Piclo Flex. Piclo Flex is an independent marketplace where DSOs and flexibility providers can trade energy to procure local flexibility services. As in Flexible Power, the products available in Piclo Flex are Sustain, Dynamic, Secure and Restore.

UK Power Networks, Scottish & Southern Electricity Networks, Western Power Distribution, SP Energy Networks and Electricity North West publish their flexibility requirements in Piclo Flex. In the case of Western Power Distribution, most of the flexibility is procured in Flexible Power, but the remaining flexibility demanded is procured in Piclo Flex [77].

Until now, products traded in Piclo Flex have a service period (time between the start and the end of use) on average of fewer than six months, arriving up to 8 years of delivery period. The bidding period is closed on average four years before the starting of the service, it is like this because a significant amount of flexibility procurement took place between 2019 and 2021 to start to be used between 2024 and 2028.

### 2.4. Collective self-consumption

Since the implementation of the European Renewable Energy Directive [78] and the Electricity Market Directive [7], EU citizens can share the energy generated by an installation with different consumers, this is known as Collective Self-Consumption (CSC). Both directives describe two roles to help the development of CSC, these are the Renewable and Citizens' Energy Communities.

The level of implementation in European countries is diverse, ranging from the impossibility of sharing the energy generated with nearby consumers to the possibility of sharing this energy with prosumers some kilometres from the point of generation.

#### 2.4.1. Spain

The Royal Decree 244/2019 [24] regulates self-consumption in Spain. It is indicated in the decree the necessity of using coefficients to distribute energy in collective self-consumption. The value of these distribution coefficients depends on the agreement between the participants, with the only requirement that they are constant values. These criteria and coefficients must be included in the agreement between the parties, which each consumer must send to the distributor directly or through its retail company. These distribution coefficients can be changed every four months (if desired) and are applied for each hour of the year for up to twenty years. If the participants do not submit different values, the regulation provides distribution coefficients calculated based on the maximum power contracted by the participating consumers.

It is also important to know that in Spain, collective self-consumption using the public grid is physically and geographically limited by the following conditions:

- Participating entities must be located within the low-voltage distribution network derived from the same transformation centre.
- The maximum distance between production and consumption meters is 500 meters. This value has been recently updated up to 2000 meters [79].
- The participating entities must be located in the same cadastral zone.

#### 2.4.2. Italy

In Italy, regulation 318/2020/R/eel [80] transposes the REDII and EMD for collective self-consumption, allowing participants in the same building or condominium to share renewable energy systems up to 200 kW.

Another option is that the prosumers can join a Renewable Energy Community (*Comunità di Energia Rinnovabile*) where production plants and withdrawal points connected to the same MV/LV substation can share energy. Residential structures, tertiary sector, industrial properties, or public administration buildings can all benefit from collective self-consumption systems.

To participate, participants must establish a private contract and name a scheme manager to represent all activities and define the generated energy distribution. The methodology for calculating the allocation of the virtual self-consumption is the result of contractual agreements between the owners and can be:

a) on an energy criterion, e.g. in proportion to the withdrawals of each user in each measurement time interval;

b) on a fixed criterion, e.g. a millesimal criterion, not related to the energy consumption of individual households.

### 2.4.3. Croatia

The Croatian law on Renewable Energy Sources and High-Efficiency Cogeneration [30] and the Electricity law define the possibility of energy sharing within a CSC framework. The information found on CSC in Croatia indicates that it is only possible to exchange energy within an energy community, the maximum power generated is 200 kW.

As in Italy, generation and consumption must be connected to the same transformation centre. There is a limit to the percentage of power that generators can cover in energy communities. The total connection power in the direction of the delivery of electricity to the network at the calculation metering points of the members of the energy community of citizens must not exceed 80% of the total connection power in the direction. Lastly, the distribution of the energy generated has to be arranged between the energy community members.

### 2.4.4. United Kingdom

Although individual self-consumption is allowed in the UK, collective self-consumption in the UK is not allowed [81], which makes this energy-sharing method impossible. Only pilot projects have been carried out in the regulatory sandbox of British regulator Ofgem "Innovation Link".

## 3. COMPARISON METHODOLOGY

In this paper, the characteristics of the countries under examination are first identified. Subsequently, an evaluation of the degree of openness of these services to the participation of prosumers is conducted. To facilitate the evaluation of the degree of openness of the countries analysed, a methodology is proposed for comparing it.

The previous review allows for the identification of similarities between the different electrical systems. One of these is the area of operation of each service, which can be divided into two main categories: national and local. These two categories serve as the foundation of the proposed methodology.

Within the national category, the common markets include the day-ahead market, intraday market, ancillary services, adequacy services, and constraint management. In the local category, the services include the local flexibility market and local energy trading.

This methodology is based on a matrix divided into different rows according to the area of operation and the market being studied, with the columns representing each development stage. To evaluate the openness of each service under examination, a score between 0 and 3 is assigned to each development stage. The significance of each score value varies depending on the service being analysed. Table 4 describes the meaning of each score value for the different services.

*Table 4 Scoring of methodology for analysing country openness to prosumer participation.*

| | SCORE | 0 | 1 | 2 | 3 |
|---|---|---|---|---|---|
| **National area** | **Day-ahead and Intraday market** | Fixed tariffs | TOU tariffs | Indexed tariffs | Explicit participation in DA and ID |
| | **Ancillary Services** — aFRR | Not allowed (+0.5 if Creation proposal) | Allowed but no aggregated / Bilateral agreements | Aggregation of large consumers | Aggregation of small consumers |
| | **Ancillary Services** — mFRR | | | | |
| | **Ancillary Services** — RR | | | | |
| | **Adequacy services** — Capacity | | | | |
| | **Constraint management services** | | | | |
| **Local area** | **Local Flexibility Market** — Long-term | Not created | Creation proposal | Aggregation of large consumers | Aggregation of small consumers |
| | **Local Flexibility Market** — Short-term | | | | |
| | **Local energy trading** | CSC not allowed | Collective self-consumption fixed | Collective self-consumption dynamic | Peer to Peer trading |
| | **Independent Aggregator?** | No | Yes | | |

In the national market, a score of 0 points is awarded if the services under examination do not permit prosumers to participate. Additional points are granted for various forms of participation, such as direct participation (1 point), participation by large consumers in aggregation (2 points), and aggregated participation of small prosumers in (3 points), with a maximum score of 3 points indicating complete market openness.

Local energy trading evaluates the option of trading between different prosumers. The first step in this process is the possibility of sharing generated energy with neighbouring prosumers (collective self-consumption). If this is feasible, a point is awarded if sharing coefficients are static (i.e. cannot be changed actively), two points are awarded if sharing coefficients can be modified dynamically, and three points are awarded if the prosumer can sell their surplus energy to anyone (peer-to-peer trading).

In the day-ahead and intraday market, points are awarded according to the type of tariffs offered, with 0 points being awarded for fixed tariffs and 2 points being awarded for indexed tariffs, with a maximum of 3 points being awarded if prosumers can participate directly in the day-ahead and intraday markets.

If the role of the independent aggregator is available for use in the country under analysis, an additional point is awarded due to this role. This is because independent aggregators help the deployment of DSF in different markets. On the other hand, traditional suppliers are inherently reluctant to offer demand flexibility programmes as these services affect their core business [82].

It is important to note that if market regulations indicate that prosumers can participate, but due to technical requirements or other reasons, they cannot actually participate, their score will be considered as 0.

The degree of openness to prosumer participation in electricity markets for a given country can be assessed by adding all the points obtained in the methodology described above. Different degrees of openness to prosumers' participation can be determined depending on the result of the sum of points of the matrix proposed. These degrees are:

- Closed market, between 0 and 6 points
- Partially open market, between 7 and 13 points
- Almost open market, between 14 and 21 points
- Open market, between 22 and 28 points

## 4. ANALYSIS OF THE COUNTRIES REVIEWED

Having established the methodology for assessing the level of openness of electricity markets to prosumer participation, a comparison is conducted among the four countries previously analysed. The scores of each country can be found in Table 5.

*Table 5 Matrix of markets' openness to participation of prosumers in Spain, Italy, Croatia and the United Kingdom.*

| | Country | | Spain | Italy | Croatia | United Kingdom |
|---|---|---|---|---|---|---|
| National area | Day-ahead and Intraday market | | 2 | 2 | 1 | 2 |
| | Ancillary Services | aFRR | 0 | 0 | 2 | 3 |
| | | mFRR | 2 | 1.5 | 2 | 3 |
| | | RR | 2 | 1.5 | 0 | 3 |
| | Adequacy services | Capacity | 0.5 | 1.5 | 0 | 1.5 |
| | Constraint management services | | 0 | 0 | 0 | 0 |
| Local area | Local Flexibility Market | Long-term | 1 | 0 | 0 | 3 |
| | | Short-term | 1 | 0 | 0 | 3 |
| | Local energy trading | | 1.5 | 1 | 1 | 0 |
| | Independent Aggregator? | | 0 | 0 | 1 | 1 |
| | TOTAL | | 10 | 7.5 | 6 | 19.5 |

Based on the methodology previously outlined, Italy has been awarded 1.5 points for mFRR and RR because these services are only available under the framework of a pilot project (UVAM) and thus cannot be awarded 2 points. Additionally, capacity markets in the United Kingdom and Italy are available for prosumer participation, but de-ratings factors in Italy and bureaucratic challenges in the UK inhibit their participation, resulting in a reduced score of 1.5 points for each country. Furthermore, in Spain, sharing coefficients in CSC can only be changed every four months, which would not be considered fully dynamic nor static, thus it has been awarded 1.5 points.

By applying the methodology defined above, it can be concluded that the United Kingdom is the most open market, with 19.5 points, and considered an almost open market. With 10 and 7.5 points respectively, Spain and Italy are classified as partially open markets. Lastly, Croatia, with a total of 2 points, is considered a closed market to prosumer participation.

# 5. CONCLUSIONS

This paper aimed to evaluate the extent of openness towards prosumer participation within four counties: Spain, Italy, Croatia, and the United Kingdom. A comprehensive examination of the various electricity markets within each country was conducted, and a methodology was proposed for comparing the 'markets' openness levels. The methodology employed in this study can be applied to analyse other European countries similarly.

The countries analysed in this study can be grouped into three categories based on their grade of openness to prosumer participation in the energy market. The United Kingdom is classified as an almost open market, with most markets open to prosumer participation and the presence of independent aggregators. However, the restriction on collective self-consumption limits its overall openness.

Spain and Italy are considered partially open markets due to limitations in the participation of small prosumers, restrictions on specific markets, and the absence of independent aggregators. These factors hinder the potential for prosumer participation and negatively impact the overall level of market openness.

Lastly, Croatia is classified as a closed market as none of its markets currently allow for prosumer participation due to technical requirements. These technical requirements pose significant barriers to prosumer participation and contribute to the overall closed status of the market.

In analysing the markets, it was found that countries within the European Union share similar characteristics due to the standardisation of markets based on European Directives. However, the UK was found to have different products and services.

Regarding participation in electricity markets, in the day-ahead and intraday markets, participation is limited to direct consumers, usually electro-intensive industries. Another form of prosumer participation in the electricity markets is through implicit demand flexibility, such as varying tariff prices to encourage the use of renewable energy or to avoid grid overload. Indexed tariffs were found in Italy, Spain, and the UK, while only Time of Use and fixed tariffs were available in Croatia.

Participation in balancing markets is already allowed in the UK and Spain, although in Spain it has only participated a demand unit for one hour in the RR market since it opened. This is probably due to the impossibility of using the role of independent aggregator, participating in balancing markets without reaching an agreement with BRPs. This is not the case in the UK because independent aggregators can participate. In Italy, prosumers cannot participate directly in the market, they have to join the UVAM pilot project and the participation has been low, so it cannot be considered fully open.

On the generation side, prosumers can self-generate energy and sell it to the grid through methods such as net metering, compensation of surplus, and payment for surplus. Italy was the most profitable self-consumption market due to net metering and direct sales to the grid. However, in other countries analysed, surplus energy prices are lower due to various factors, including transmission and distribution costs in Spain, and pricing determined by retailers or laws in the UK and Croatia.

Collective self-consumption was not allowed in all countries analysed, presenting a barrier for prosumers who may not have the resources or desire to generate energy individually, limiting the implementation of distributed renewable technologies.

Regarding local flexibility markets, it is only available now in the United Kingdom, in Piclo Flex and Flexible Power platforms. A proposal since 2019 has been developed in Spain, known as IREMEL, and pilot projects are currently in process. The creation of such markets has not been envisaged in Croatia and Italy.

Despite the presence of European Directives, barriers such as the inability to use the independent aggregator or minimum power requirements make it challenging for prosumers to participate in the markets. In Spain, Italy and especially Croatia, efforts must be made to allow prosumers to participate in all markets.

## ACKNOWLEDGEMENTS

This paper has been prepared in the framework of the ReDREAM project, funded by the European Union's Horizon 2020 research and innovation programme (Horizon2020 Framework Programme) under grant agreement number 957837.